# Load balancing in cloud data centers with optimized virtual machines placement

Hamid Reza Naji
*Dept. Computer Engineering and Information Technology*
*Graduate University of Advanced Technology*
Kerman, Iran
naji@kgut.ac.ir

Reza Esmaeili
*Dept. Computer Engineering and Information Technology*
*Graduate University of Advanced Technology*
Kerman, Iran
r.esmaeili@kgut.ac.ir

*Abstract*— **So far, various solutions have been proposed for symmetric distribution of load cloud computing environments. In this article, a new solution to the optimal allocation of virtual machines in the cloud data centers is presented to provide a good load balancing among servers. The proposed method offers a solution uses learning automata as a reinforcement learning model to improve the performance of the optimization algorithm for optimal placement of virtual machines. Also, it helps the search algorithm to converge more quickly to the global optimum. The simulation results show the proposed method has been able to perform good level of load balancing in cloud data centers.**

## I. Introduction

In a data center, most services often need only a small part of the total available resources. In order to avoid resource waste, several virtual machines are placed over the least possible number of physical servers and the rest of the inessential servers go to sleep mode (low power consumption mode). This approach known as "server consolidation", can prevent overconsumption of servers. Decreasing the number of servers along with improvement and complexity reduction of infrastructure in addition to cost and energy savings have remarkable advantages for data centers [1,2,3,4,5]. In continue, cloud computing and virtualization are discussed briefly and also the research background is presented. The proposed method and its simulation results are provided. Finally, the conclusion and future works in mentioned.

## II. Literature Review

Owing to the demand for computational power that is needed for application programs, information technology infrastructures have been growing rapidly in recent years and modern data centers using cloud computing are hosting various advanced application programs. But at the same time, high energy costs and greenhouse gas emissions are serious problems resulting from utilization of big data centers. In the last few decades, a huge progress is done in the production of data storage systems, processing systems and network technologies to share these two and other computing resources to develop cloud computing environments. Actually, in a cloud environment computing resources are presented to the users as an internet service. Here, you do not own the presented service and merely use it [6].

Many big companies working in the field of internet, have developed hardware/software infrastructure since 2000 in order to meet their processing needs and also protect their data centers from physical and cyber-attacks [7]. Generally speaking, virtualization is a basis for the resource integration of independent systems so that through this technology, equipment and resources can be used longer and consequently, the rapid return of capital will be assured [8].

Ensuring service consistency is a challenge in regard to virtualization use, especially during the reconfiguration and relocation of a large portion of sellers and their implementation in various hypervisors [9]. Virtualization infrastructures must be able to develop platforms in appropriate locations and time so that they are provided with allocated dynamics and hardware resource scalability [10].

There's a wide range of participation and innovations to "green up" cloud data centers. An interesting investigation has been presented in reference [11], which provides concept apprehension and the need for provision of energy-aware resources. In reference [12], methods for the virtual machine placement, server consolidation and load balancing have been mentioned and investigated in order to detect their faults and cause and effect relationships. Similarly, reference [13] discusses the open challenges in dynamic resource allocation. Multiple discoveries concerning virtual machine placement have been compared and updated in reference [14]. The study conducted in reference [15], generally classifies virtual machine placement methods into two categories, direct placement and placement based on migration, which is based on the duration of time needed for a work completion depending on the type of virtual machine placement. The impact of virtual machine placement performance and competition over resources have been argued in reference [16], while in reference [17], despite the energy saving approaches mentioned above, researchers have tried to maximize the placement ratio, because their main concentration has been on the compact placement.

## III. THE PROPOSED METHOD

The form of virtual machine allocation to network physical machines is the main issue in this architecture. Every virtual machine has a set of requirements. The allocation of resources to virtual machines should be in a way that any hardware can meet all needs of virtual machines. On the other hand, the allocation pattern of virtual machines directly affects parameters such as load balancing, energy consumption and cloud throughput. In this paper a new solution to the optimal allocation of virtual machines in the cloud computing system will be presented to provide a load balanced cloud data center. In the following, the proposed method will be presented after explaining the system model. In our model we use a cloud controller which is explained in the following:

### A. Cloud Controller:

This component is in charge of configuring computing resources in the cloud structure. The Cloud Controller stores and restores the operational status of the service providers available on the network and provides the service providers' capacity for the Resource Optimizer (for instance: the processor, the memory and etc.). Additionally, the Cloud Controller receives the optimal solution found by the proposed allocation algorithm from VM Orchestrator to detect and properly allocate the service providers' resources to virtual machines by using a proper API such as Cloud Stack.

As mentioned earlier, in this architecture the multi-objective cuckoo search optimization algorithm is used for optimal allocation of virtual machines in the cloud structure. Therefore, the optimization algorithm is implemented in the Resource Optimizer component and the found solution is delivered to the VM Orchestrator and it delivers the found solution to the VM Orchestrator). Following this chapter, the way of allocating resources to virtual machines by using the proposed method will be discussed.

### B. Optimal Allocation of Virtual Machines

In any problem of virtual machine allocation, there might be a lot of reliable solutions. However, determining the optimality of each of the existing reliable answers is the main issue in such a case/situation. It should be possible to determine the best possible status for the problem of virtual machine allocation by using an accurate and efficient criterion. In this problem, it should be taken into consideration that targeting a parameter or employing an objective parameter may not lead to desirable results to solve the research problem. That's why, in the proposed method the multi-objective optimization algorithm is used to solve the virtual machine allocation problem. MOCS is the algorithm used in the present study.

However, the search history of the optimization algorithm can be utilized in order to create new answers. To this purpose, the learning automaton and the reinforcement learning approach will be employed in order to develop random answers which are close to the global optimum. The combination of the multi-objective cuckoo search optimization algorithm and the learning automaton for faster detection of the global optimal answer is one of the innovations mentioned in the proposed method. Simultaneous consideration of the criteria such as energy consumption, load balancing and resource utilization rate as optimization objectives, is also recognized as the second innovative aspect of the current study. The combination of these two approaches can be effective in converting the proposed algorithm into an optimal mechanism for the virtual machine placement. The procedure will be discussed in the following.

Every learning automaton is defined by a set of selectable actions. In the following, the set of actions will be represented by A= $\{\alpha_1, \alpha_2, \ldots, \alpha_n\}$. Each selectable action in A has a probability value and the selection of each action is made based on this set of probabilities. Selecting a member of action set A and exerting it on the environment, the learning automaton starts its action. The exerted action will be evaluated by the environment and the automaton selects its next action based on the response received from the environment. During each action selection, in case the received response from the environment is desirable, the learning automaton increases the probability corresponding to the selected action and in case of receiving an undesirable response from the environment, the probability of that action decreases.

Thus, a learning automaton is defined for each virtual machine in which each learning automaton has **M** selectable actions (**M** is the number of physical machines). The aim of each automata model is to present an approximation of the selectable physical servers for the virtual machine corresponding to it. At the first iteration of the multi-objective cuckoo search optimization algorithm, all actions have the same probability of $\left(\frac{1}{M}\right)$. Considering this structure based on the learning automata, the optimization algorithm creates half of the responses based on the learning automata's probabilities (instead of randomization of a response vector structure in the problem space). The reason for determining half of the responses based on the learning automata is that the full search probability in the problem space can still be maintained, thus preventing the optimization algorithm from the probability of sticking in the local minimum. Therefore, the pseudocode of the proposed algorithm for the optimal placement of virtual machines based on the combination of the multi-objective cuckoo search optimization algorithm and the learning automata will be as follows:

At the end of each cycle, the worst and the best responses of the population will be utilized to update the learning automata's structures. Therefore, the best and the worst responses in the current population will be detected first. Then the probability of all actions corresponding to the best response of the current population in learning automata models is increased by the reward operator. This operation is performed for every learning automata model using the relation (1).

$$p_j(k+1) = \begin{cases} p_j(k) + a[1 - p_j(k)] & j = i, \\ (1-a)p_j(k) & \forall j \neq i. \end{cases} \quad (1)$$

In the above relation, **a** is the reward factor which is considered equal to 5.0. $p_j(k)$ indicates the selection probability of the **j**th action of learning automaton in the **k**th repetition. For instance, if the best response in the current population has placed the virtual machine **x** in the physical server **w**, the probability of action **w** in the learning automaton corresponding to virtual machine **x** will be updated by the relation 1.

Considering this searching mechanism, the response vector structure, the optimization objectives and the problem limitations will be analyzed in the following.

### C. Response vector structure and optimization objectives

In this chapter, the response vector structure and the target criteria for fitness evaluation in the proposed algorithm will be explained. In this stage, the objective of the optimization algorithm is the optimal allocation of virtual machines to physical machines. If the problem under discussion has **n** virtual machines and **m** physical machines, each response vector in the optimization algorithm will have a length equal to **n**. Each entry of this vector indicates a virtual machine and the number existing in this entry shows the ID of the service provider to which the virtual machine has been allocated in the current answer. Therefore, each entry of the solution vector can have one of the quantities from 1 to **m**. In other words, the number of optimization variables is equal to **n** and the bounds of each optimization variable will be in the form of [1, **m**].

### D. Utilization rate of physical resources

The aim of a virtual machine allocation algorithm is the maximum resource utilization of network physical systems. This criterion is calculated as allocated physical resources to the total available resources:

$$maximize\ U = \sum_{j=1}^{m}\frac{\sum_{i=1}^{n}(x_{ij}.\alpha R_{p_i})}{T_{p_j}} + \sum_{j=1}^{m}\frac{\sum_{i=1}^{n}(x_{ij}.\beta R_{m_i})}{T_{m_j}} \quad (2)$$

In the above relation, $R_{p_i}$ indicates the amount of CPU required in each virtual machine such as **i** and $T_{P_j}$ indicates the maximum processing power in each server such as **j**. $R_{m_i}$ also shows the amount of main memory required in the virtual machine **i** and $T_{m_j}$ shows the total amount of main memory in the service provider **j**. In the above relation, **α** is the processing resource value required by the client and **β** is the memory resource value required by the client. $x_{ij}$ is the descriptive binary function of the virtual machine **i** allocation to service provider **j**, so that if virtual machine **i** locates in service provider **j**, the function value of $x_{ij}$ will be equal to one and otherwise it will be equal to zero.

### E. Load balancing

This objective controls the balanced distribution of computational load among physical servers and employing this criterion as one of the optimization objectives can assure the computational performance of the cloud computing system in the virtualization process. In order to calculate the load balancing, standard deviation concept can be used and the utilization rate of physical resources can be balanced through the relation below:

$$minimize\ LB = \sqrt{\frac{\sum_{i=1}^{m}(U_i-\bar{U})^2}{m}} \quad (3)$$

In the relation above, **m** is the number of physical machines and $U_i$ is the utilization rate of **i**th physical resource which is computable via relation 1. $\bar{U}$ is also the average utilization rate of all active physical resources in the response detected by the optimization algorithm.

### F. restrictions of the proposed algorithm

Restrictions are equations or inequalities which are placed next to the objective function and represent limitations of each of the problem variables. We're facing two restrictions in the problem of network virtual machine allocation through the proposed method. Formulating each of these limitations will be discussed in the following.

The purpose of exerting this limitation is the operationally of a detected solution by the proposed method in a real situation. Accordingly, a service provider can't implement virtual machines whose total demand volume is greater than the service provider processing power. Therefore, if the total demand of the virtual machines allocated to the server **j** is equal to $D_j$ and if the server **j** has the capacity of $C_j$.

## IV. THE IMPLEMENTATION OF THE PROPOSED METHOD

The proposed algorithm has been implemented by MATLAB software and the proposed algorithm is studied from load balancing and processing time aspects. The results of the simulation for the proposed method is compared with the previous methods.

### A. Simulation

A random database consisting of some hypothetical virtual and physical machines has been used in order to evaluate the proposed method. Every virtual machine has some requirements and has implementation ability in the physical machine so that it can meet the process requirements. It is also supposed that all physical machines have the possibility of parallelization. Therefore, every service provider can host more than one virtual machine. Thus, every input problem in the simulation process has been developed by hypothetical data. That's why, at first, **N** number of physical machines with random processing power have been defined in specific intervals. The central processor power of every server has been described as a random number in the interval gigahertz and the amount of main memory has been defined as a random number in the interval gigabytes. After defining the physical machines, the number of N<K virtual machines with random requirements and in specific intervals have been developed. These virtual machines have been defined in a way that their total requirements in every problem would be at least 90% of the total servers' processing power. Additionally, the requirements of every virtual machine processor and main memory have been defined less than average amount of the ability of these specifications for physical machines. Thus, in the present problem, at least one physical machine corresponding to the requirements of each virtual machine can be definitely found.

During the simulation operation, results of the proposed method for the problems of virtual machine allocation are investigated from different aspects. That is, by changing the number of virtual machines available in the problem, the allocation operation for these machines will be done through the proposed method. At the end of the simulation, the following criteria have been detected to evaluate the results of the proposed method:

Load balancing in virtual machine allocation: An efficient allocation algorithm, distributes the computational load among the physical servers in a balanced way and prevents imposing the computational load on a specific service provider. In the experiments conducted, in order to calculate the load balancing, the standard deviation concept has been used and the utilization rate of physical resources has been balanced through previous relations. In the following, the results achieved from the proposed method implementation will be explained and its throughput will be compared to and contrasted with the previous algorithms.

## V. SIMULATION RESULTS

Simulation operation has been performed based on the mentioned parameters and scenarios. During conducting the experiments, the number of virtual machines has been changed from 20 to 100 and the six criteria i.e. energy consumption, resource waste, number of the service providers employed, rate of service provider's CPU utilization, load balancing and processing time have been explored. Moreover, the results achieved from the proposed method evaluation have been compared to and contrasted with GABP algorithm in the reference [18] and PSO algorithm in the reference [19]. The method presented in the reference [18], uses Genetic Algorithm and the method presented in the reference [19], uses Particle Swarm Optimization algorithm in order to solve the problem of virtual machine placement in cloud computing resources. The parameters employed in the proposed algorithm are as follows: population size: 100, maximum number of cycles: 500, $P_a$ probability: 0.25

Furthermore, in the compared algorithm (Genetic), the population size and the number of generations have been respectively determined equal to 100 chromosomes and 500 cycles. The combination rate and the mutation rate have been considered equal to 0.7 and 0.05 respectively. In Particle Swarm Algorithm, the parameters of the population size and the number of repetitions have been also considered the same as the above values. It is worth mentioning that in all the experiments conducted in this chapter, the simulation process has been repeated 10 times and the average of the results achieved from the 10-time independent repetition of simulation, have been presented.

We have named our proposed method as LAMOCS (Learning Automata-based Multi-Objective Cuckoo Search) so in simulation diagrams you can recognize it in compare to other methods. Fig. 1 shows the diagram of the average number of servers employed in the optimal response of the allocation process based on the number of virtual machines. Utilizing fewer servers is the acceptable state for this criterion. In Fig. 1, the number of virtual machines available in the problem is shown on the horizontal axis and the number of servers employed for allocation in the optimal response is shown on the vertical axis.

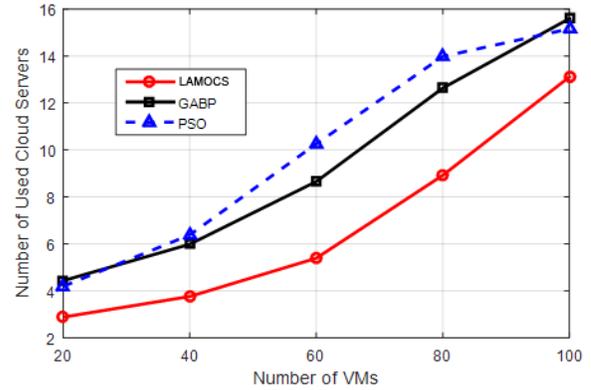

fig. 1. The average number of servers based on the number of virtual machines

According to the results illustrated in Fig. 1, through combining the multi-objective cuckoo search optimization algorithm and learning automaton, the proposed approach can perform the allocation operation with fewer servers. That's why one of the objectives of the fitness employed in the proposed method is to implement the allocation operation with the minimum amount of energy consumption. This characteristic also refers to the number of servers indirectly. Because an increase in the number of service providers leads to an increase in the energy consumption of the service providers on standby mode.

In accordance with the achieved results, the proposed algorithm outperforms the compared algorithms and on the average, it can employ fewer servers in the allocation process.

Fig. 2 shows the diagram of the average amount of the central processor usage in service providers side based on the number of virtual machines. In this diagram, the number of virtual machines is demonstrated on the horizontal axis and the vertical axis demonstrates what percent of the processor capacity has been utilized on the average, in each employed server.

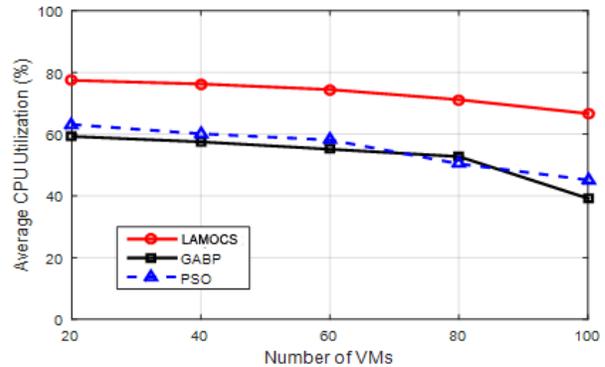

Fig. 2 The average amount of the processor usage in servers based on the number of virtual machines.

Although the results presented so far reveal the method performance in the optimization of virtualization architecture in a cloud computing system, confirming these results depends on the quality of computational load distribution among service providers. As a result, the load balancing criterion must be studied as one of the significant criteria in the performance evaluation of cloud computing systems. These results are illustrated in Fig. 3. In Fig. 3, in order to describe the amount of load balancing in the cloud, the concept of standard deviation has been used to calculate the amount of computing resource utilization through previous relations.

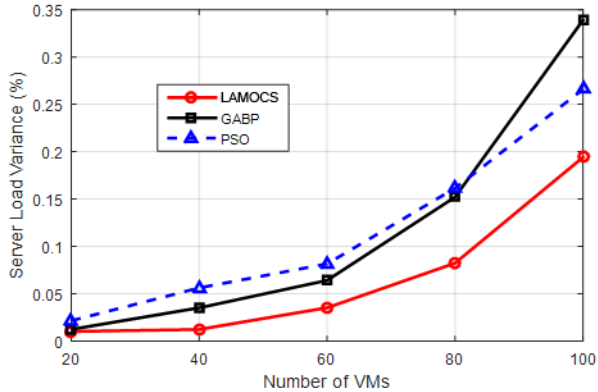

Fig. 3. The servers' load balance based on the number of virtual machines.

A mechanism of virtual machine allocation to cloud physical resources will achieve load balancing on condition that the utilization amount of resources enabled in the cloud is almost uniform and in other words, the standard deviation of the utilization amount of enabled computing resources is minimized. In Fig. 3, this condition is thoroughly observable for the method. Considering the load balancing criterion as one of the optimization objectives in the method has caused the proposed approach to be able to distribute the computational load resulting from virtual machines more uniformly among cloud physical resources in different situations.

Fig. 4, shows the diagram of processing time of virtual machine allocation algorithm for changes in the problem dimensions. This diagram shows how many seconds on average, each of the compared algorithms spends to search for the final response. This experiment has been implemented on a desktop computer with intel i7 processor and 1.8 Gigahertz of processing frequency and 8 Gigabytes of main memory.

Fig. 4 demonstrates that as the number of virtual machines increases, the time required for the implementation of the optimization algorithm increases. Because as previously mentioned, an increase in the number of virtual machines leads to an increase in the problem dimensions and the solution vector length. Therefore, the optimization algorithm must implement more computational operations.

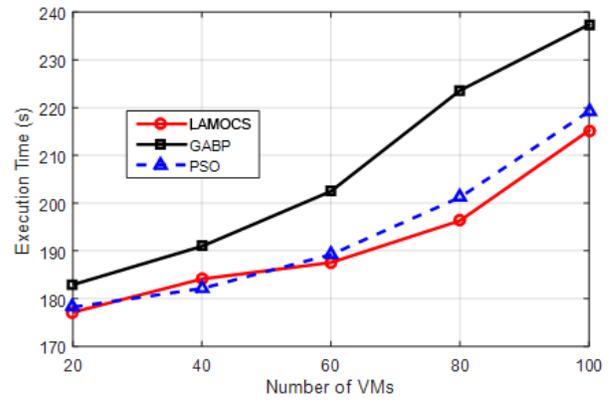

Fig. 4. The processing time of virtual machine allocation algorithm based on the number of virtual machines

However, according to the results revealed in Fig. 4, the method spends less time on searching for the optimal response. That's why in the GABP method [18], Genetic algorithm has been utilized to solve the problem and regarding the multiple operators employed in this algorithm, it needs more computational operations than cuckoo search algorithm does. On the other hand, Particle Swarm Optimization algorithm in reference [18], because of its fewer computational operators than the Genetic algorithm's, has the same implementation time as the method does. Besides, in Fig. 5, the diagram of the processing time of virtual machine allocation algorithms for changes in the population size is demonstrated. In this diagram, the number of virtual machines has been considered 100.

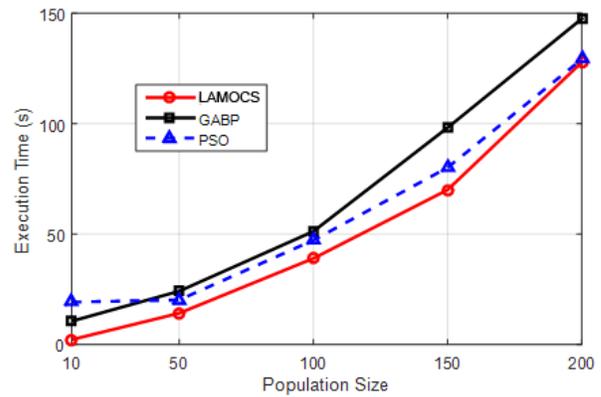

Fig. 5. The processing time of virtual machine allocation algorithm based on the population size

By employing operators with more computational complexity, the Genetic algorithm has more total time than Cuckoo Search algorithm does; and in respect of the implementation time, the Particle Swarm Optimization algorithm, has the same throughput as the method does.

Based on the experiments conducted in the present research, our method can perform optimization operation in a shorter period of time and achieve more optimal states of load balancing.

## VI. CONCLUSION AND SUGGESTIONS

### A. Conclusion

In the present study, a new method to the optimal allocation of virtual machines in the cloud computing system has been presented. The method uses load balancing criteria in order to evaluate the fitness of each answer. In this method, the answer with the least rate of the cloud resource waste, is chosen as the final solution. MATLAB software has been utilized to implement the method. Moreover, in the simulation procedure, problems with different sizes (different number of virtual machines) have been used. The results of the experiments conducted, indicate that the answers created by the method for virtual machine allocation provides a high level of load balancing of cloud resources. According to the results achieved from the experiments, considering the load balancing criterion as one of the optimization objectives in the method has caused the proposed approach to be able to distribute the computational load, resulting from virtual machines, more uniformly among cloud physical resources in different situations. Furthermore, the method, has higher processing speed than the compared algorithm does so that it can perform the processing operation with a higher speed than the other methods.

### B. Suggestions

More suitable results may be achieved through changing the fitness objectives in the multi-objective optimization algorithm. Employing other multi-objective optimization algorithms such as multi-objective Particle Swarm, multi-objective Genetic and etc. for the optimal allocation of virtual machines to cloud physical resources can be investigated in future studies.